\documentclass{amsart}
\usepackage{amsmath}
\usepackage{amsfonts,amssymb}

\usepackage{amsmath,amsthm,url,graphicx, latexsym, curves}
\usepackage{amsfonts}
\usepackage{amssymb}

\providecommand{\U}[1]{\protect\rule{.1in}{.1in}}

\newtheorem{theorem}{Theorem}
\newtheorem{lemma}[theorem]{Lemma}
\newtheorem{proposition}[theorem]{Proposition}

\theoremstyle{remark}
\newtheorem{remark}[theorem]{Remark}

 \numberwithin{equation}{section}

\begin{document}

\author[ Dang-Zheng Liu  and Da-Sheng Zhou ]{ Dang-Zheng Liu and Da-Sheng Zhou}

\address{School of Mathematical Sciences, Peking University, Beijing,
100871, P.R. China} \email{dzliumath@gmail.com}
\address{Department of Mathematics, University of Macau, Av. Padre Tom\'{a}s
Pereira, Taipa, Macau, P.R. China} \email{zhdasheng@gmail.com}

\title [Universal properties for restricted  Trace Gaussian Ensembles]{some
Universal properties for restricted  Trace Gaussian orthogonal, unitary and Symplectic
 Ensembles}
%\author{Dang-Zheng Liu\inst{1}\and Da-Sheng Zhou\inst{2}}
%%\thanks{\emph{E-mail :} \inst{1}Corresponding author: dzliumath@gmail.com; \inst{2} zhdasheng@gmail.com}}
%\institute{School of Mathematical Sciences, Peking University,
%Beijing, 100871, P.R. China.\\E-mail: dzliumath@gmail.com
%\\ \and Department of Mathematics,
%University of Macau, Av. Padre Tom\'{a}s Pereira, Taipa, Macau, P.R.
%China.\\
%E-mail: zhdasheng@gmail.com}

%\maketitle
%
% Add name of the expert who has communicated your paper
%\communicated{name}
%
\maketitle
\begin{abstract}
Consider  fixed and bounded trace Gaussian  orthogonal, unitary and
symplectic  ensembles, closely related to Gaussian ensembles without
any constraint. For three restricted trace Gaussian ensembles, we
prove universal limits of correlation functions at zero and at the
edge of the spectrum edge. In addition, by using  the universal
result in the bulk  for fixed trace Gaussian   unitary  ensemble,
which has been obtained by G$\ddot{o}$tze and  Gordin, we also prove
the universal limits of correlation functions for bounded trace Gaussian
unitary ensemble.
\end{abstract}
\section{Introduction and Main Results}
\label{intro}

The standard random matrix models are Gaussian orthogonal, unitary
and symplectic ensembles, denoted respectively by GOE, GUE and GSE.
The joint probability density function (p.d.f.) for the eigenvalues
$x_{1},\ldots,x_{N}$ of  these three canonical ensembles is given
\cite{mehta} by
\begin{equation} \label{pdfGbeta}
P_{N\beta}(x_{1},\cdots,x_{N})=\frac{1}{Z_{N\beta}}\prod_{1\leq
j<k\leq N}|x_{j}-x_{k}|^{\beta}\prod_{i=1}^{N}e^{-\beta Nx_{i}^{2}}
\end{equation}
where the partition function
\begin{equation}
\label{partition function} Z_{N\beta}=(2\pi)^{\frac{N}{2}}(2\beta
N)^{-\frac{N_\beta}{2}}\prod_{j=1}^{N}\frac{\Gamma(1+\frac{\beta
j}{2})}{\Gamma(1+\frac{\beta}{2})}.
\end{equation}
Here $N_{\beta}= N+\beta N(N-1)/2$, and $\beta=1, 2$ or 4 correspond
to  GOE, GUE or GSE respectively.
%$N_{\beta}=N+\frac{\beta N(N-1)}{2}$.
The $n$-point correlation functions for the Gaussian ensembles are
defined \cite{dyson,mehta} by
\begin{equation}
\label{correlation function}
R_{n\beta}(x_{1},\ldots,x_{n})=\frac{N!}{(N-n)!}\int_{\mathbb{R}^{N-n}}
P_{N\beta}(x_{1},\ldots,x_{N})dx_{n+1}\cdots dx_{N},
\end{equation}
which measures the probability density  of finding a level
(regardless of labeling) around each of the points
$x_{1},\ldots,x_{n}$, the positions of the remaining levels being
unobserved. In particular, $R_{1\beta}$  gives the overall level
density. A classic result  shows that the asymptotic normalization
level density $\frac{1}{N}R_{1\beta}$ (density of states) as
$N\rightarrow \infty$ is given by the Wigner semicircle law
$\omega(x)=\frac{2}{\pi}\sqrt{(1-x^2)_{+}}$ for  $\beta=1,2,4$.
 However, when
$\beta=1,2,4$, for $n\geq 2$, the study of a finer asymptotics
presents a pattern (called universality) in the bulk $(-1,1)$ and at
the edge $\pm 1$ of the spectrum. %To present this universal pattern,

To present this universal pattern, let us introduce a so-called
integral kernel \cite{tw2,soshnikov1,soshnikov2}
\begin{equation}
K(x,y)=\frac{\varphi(x)\
\varphi^{'}(y)-\varphi^{'}(x)\,\varphi(y)}{x-y}\label{Kxy}
\end{equation}
where $\varphi(x)$ is a real-valued function. When one takes
$\varphi(x)=\frac{\textrm{sin}\, \pi x}{\pi}$ or  $\textrm{Ai} (x)$,
$K(x,y)$ in $(\ref{Kxy})$ will be rewritten as
$K_{\textrm{sine}}(x,y)$ and $K_{\textrm{Airy}}(x,y)$, respectively.
Here $\textrm{Ai} (x)$ stands for the Airy function satisfying the
differential equation
$$\textrm{Ai}''(x)=x\,\textrm{Ai}
(x).$$ Universality of sine-kernel in the bulk for GOE, GUE and GSE
\cite{Brezin, mehta,tw3} says: for every $u \in (-1,1)$ and
$t_{1},\ldots,t_{n} \in \mathbb{R}^{1}$,
\begin{equation}
\lim_{N \rightarrow \infty}\label{sinekernel}
\frac{1}{(N\omega(u))^{n}}
R_{n\beta}(u+\frac{t_{1}}{N\omega(u)},\ldots,u+\frac{t_{n}}{N\omega(u)})=\begin{cases}
\textrm{det}[K_{\textrm{sine}}(t_{j},t_{k})]_{j,k=1}^{n},\ \ \ \ \  \beta=2\\
\big(\textrm{det}[K_{\textrm{sine}}^{(\beta)}(t_{j},t_{k})]_{j,k=1}^{n}\big)^{\frac{1}{2}},
\ \beta=1,4,
\end{cases}
\end{equation}
where
\begin{equation}
\label{K1} K_{\textrm{sine}}^{(1)}(x,y)=\begin{pmatrix}
  K_{\textrm{sine}}(x-y) & \frac{\partial}{\partial x}K_{\textrm{sine}}(x-y)   \\
\int_{0}^{x-y}K_{\textrm{sine}}(t)dt-\frac{1}{2}\textrm{sgn}(x-y) &
K_{\textrm{sine}}(x-y)
\end{pmatrix},\end{equation}
and
\begin{equation}
\label{K4} K_{\textrm{sine}}^{(4)}(x,y)=\begin{pmatrix}
  K_{\textrm{sine}}(2(x-y)) & \frac{\partial}{\partial x}K_{\textrm{sine}}(2(x-y))   \\
\int_{0}^{x-y}K_{\textrm{sine}}(2t)dt-\frac{1}{2}\textrm{sgn}(2(x-y))
& K_{\textrm{sine}}(2(x-y))
\end{pmatrix}.\end{equation}
It has been proved that this is also true in other invariant
ensembles under the orthogonal, unitary or symplectic groups
\cite{ps,bi,dkmvz,dg1}  and in certain ensembles of Hermitian Wigner
matrices \cite{johansson2}. Similarly, universality of Airy-kernel
at the edge of spectrum for GOE, GUE and GSE
\cite{forrester1,tw1,mehta} says: for $t_{1},\ldots,t_{n} \in
\mathbb{R}^{1}$,
\begin{equation}
\lim_{N \rightarrow \infty}\label{airykernel} \frac{1}{(2
N^{\frac{2}{3}})^{n}} R_{n\beta}(1+\frac{t_{1}}{2
N^{\frac{2}{3}}},\ldots,1+\frac{t_{n}}{2
N^{\frac{2}{3}}})=\begin{cases}
\textrm{det}[K_{\textrm{Airy}}(t_{j},t_{k})]_{j,k=1}^{n},\ \ \ \ \  \beta=2\\
\big(\textrm{det}[K_{\textrm{Airy}}^{(\beta)}(t_{j},t_{k})]_{j,k=1}^{n}\big)^{\frac{1}{2}},
\ \beta=1,4.
\end{cases}
\end{equation}
where
\begin{equation}
\label{K3} K_{\textrm{Airy}}^{(1)}=\begin{pmatrix}
  (K_{\textrm{Airy}}^{(1)})_{11} & (K_{\textrm{Airy}}^{(1)})_{12}   \\
(K_{\textrm{Airy}}^{(1)})_{21} & (K_{\textrm{Airy}}^{(1)})_{22}
\end{pmatrix},\end{equation}
\[(K_{\textrm{Airy}}^{(1)})_{11}=(K_{\textrm{Airy}}^{(1)})_{22}=K_{\textrm{Airy}}(x,\,y)+\frac{1}{2}\,\textrm{Ai}(x)\left(1-\int_{y}^{\infty} \textrm{Ai}(z)\,dz\right),\]
\[(K_{\textrm{Airy}}^{(1)})_{12}=-\partial_{y}K_{\textrm{Airy}}(x,\,y)-\frac{1}{2}\textrm{Ai}(x)\textrm{Ai}(y),\]
\[(K_{\textrm{Airy}}^{(1)})_{21}=-\int_{x}^{\infty} K_{\textrm{Airy}}(z,y)\,dz\]
\[+\frac{1}{2}\left(\int_{y}^{x} \textrm{Ai}(z)\,dz+
\int_{x}^{\infty} \textrm{Ai}(z)\,dz\,\cdot\,\int_{y}^{\infty}
\textrm{Ai}(z)\,dz\right)-\frac{1}{2}\textrm{sgn}(x-y),\] and in the
case $\beta=4$,
\begin{equation}
\label{K4} K_{\textrm{Airy}}^{(4)}=\frac{1}{2}\begin{pmatrix}
  (K_{\textrm{Airy}}^{(4)})_{11} & (K_{\textrm{Airy}}^{(4)})_{12}   \\
(K_{\textrm{Airy}}^{(4)})_{21} & (K_{\textrm{Airy}}^{(4)})_{22}
\end{pmatrix},\end{equation}
\[(K_{\textrm{Airy}}^{(4)})_{11}=(K_{\textrm{Airy}}^{(4)})_{22}=K_{\textrm{Airy}}(x,\,y)-\frac{1}{2}\,\textrm{Ai}(x)\int_{y}^{\infty} \textrm{Ai}(z)\,dz,\]
\[(K_{\textrm{Airy}}^{(4)})_{12}=-\partial_{y}K_{\textrm{Airy}}(x,\,y)-\frac{1}{2}\textrm{Ai}(x)\textrm{Ai}(y),\]
\[(K_{\textrm{Airy}}^{(4)})_{21} =-\int_{x}^{\infty} K_{\textrm{Airy}}(z,y)\,dz+\frac{1}{2}\int_{x}^{\infty}\textrm{Ai}(z)dz\int_{y}^{\infty}\textrm{Ai}(z)dz.\]
This Airy-kernel has been proved true in other invariant ensembles
\cite{bi,dg2}, and in real symmetric and Hermitian Wigner random
matrices \cite{soshnikov1}. Notice that we have used a quaternion
determinant \cite{mehta} defined by Dyson,  after
$K_{\textrm{sine}}^{(1)}$, $K_{\textrm{sine}}^{(4)}$,
$K_{\textrm{Airy}}^{(1)}$, $K_{\textrm{Airy}}^{(4)}$, these
 $2\times 2$ matrices  are  thought of as quaternions.

In the present paper, we will deal with three restricted trace
Gaussian
 ensembles: fixed  trace Gaussian
 ensembles and bounded  trace Gaussian
 ensembles. The aim is to extend the properties (\ref{airykernel}) and  (\ref{sinekernel}) when $u=0$ to the fixed
and bounded trace GOE, GUE and GSE. First, let us give a review
\cite{rosenzweig,mehta} for restricted trace ensembles. Proceeding
from the analogy of a fixed energy in classical statistical
mechanics, Rosenzweig defines \cite{rosenzweig} his ``fixed trace''
ensemble for a Gaussian real symmetric, Hermitian or self-dual
matrix $H$ by the requirement that the trace of $H^{2}$ be fixed to
a number $r^{2}$ with no other constraint. The number $r$ is called
the strength of the ensemble. The joint probability density function
for the matrix elements of $H$ is therefore given by

\[P_{r}(H)=K_{r}^{-1}\,
\delta\big(\frac{1}{r^{2}}\text{tr}\,H^{2}-1\big)\]
 where $K_{r}$ is the normalization constant.
  Note that this probability
density
 function is invariant under a conjugate action by orthogonal, unitary
 or symplectic groups, because of the invariance of the quantity
 $\text{tr}\,H^{2}$. Its eigenvalue joint p.d.f. has the form

\begin{equation}
\label{fixedtracepdf}
P_{N\beta}^{FT,r}(x_{1},\cdots,x_{N})=\frac{1}{Z_{N\beta}^{FT,r}}\delta
(\sum_{i=1}^{N}x_{i}^{2}-r^{2}) \prod_{1\leq j<k \leq N}
|x_{j}-x_{k}|^{\beta},
\end{equation}
where the normalization constant $Z_{N\beta}^{FT,r}$ can be computed
by virtue of variable substitution for the partition function
 $Z_{N\beta}$ \cite{ld}:
\begin{equation}
\label{partition function for fixed trace}
Z_{N\beta}^{FT,r}=r^{N_{\beta}-1}\frac{(2\pi)^{\frac{N}{2}}2^{-\frac{N_{\beta}}{2}+1}}{\Gamma(\frac{N_{\beta}}{2})}\prod_{j=1}^{N}\frac{\Gamma(1+\frac{\beta
j}{2})}{\Gamma(1+\frac{\beta}{2})}.
\end{equation}
Here $N_{\beta}=N+\beta N(N-1)/2$. Notice the analogy: fixed trace
ensemble  bears  the same relationship to the unconstrained ensemble
that the microcanonical ensemble to the canonical ensemble in
statistical physics \cite{balian}. Instead of keeping the trace
constant we might require it to be bounded \cite{bronk,mehta},  the
eigenvalue joint p.d.f for the bounded trace Gaussian ensembles is
given by
\begin{equation}
\label{boundedtracepdf}
P_{N\beta}^{BT,r}(x_{1},\cdots,x_{N})=\frac{1}{Z_{N\beta}^{BT,r}}\theta
(r^{2}-\sum_{i=1}^{N}x_{i}^{2}) \prod_{1\leq j<k \leq N}
|x_{j}-x_{k}|^{\beta},
\end{equation}where $\theta (x)$ denotes the Heaviside step function, i.e.,
$\theta (x)=1$ for $x\geq 0$, otherwise $\theta (x)=0$.
 Proceeding
further, G. Akemann et al \cite{acmv,av} have considered a
generalization of a fixed or bounded ensemble up to an arbitrary
polynomial potential, and described further interesting physical
features of restricted trace ensembles due to the interaction among
eigenvalues introduced through a constraint. In addition, we also
notice that fixed and bounded trace ensembles are norm-dependent
ensembles. Recently, T. Guhr \cite{Guhr1,Guhr} has  obtained the
supersymmetric integral representations of the more general
correlation function from norm dependent random matrix ensembles to
arbitrary unitarily invariant matrix ensembles by generalizing the
Hubbard--Stratonovich transformation.

As done usually for Gaussian ensembles in Eq. (\ref{correlation
function}), the $n$-point correlation function for fixed trace
ensembles is defined
 by

\begin{equation}
\label{correlation function for fixed trace}
R_{n\beta}^{FT,r}(x_{1},\ldots,x_{n})=
\frac{N!}{(N-n)!}\int_{\mathbb{R}^{N-n}}
P_{N\beta}^{FT,r}(x_{1},\ldots,x_{N})dx_{n+1}\cdots dx_{N}.
\end{equation}
More precisely,
\begin{equation}
R_{n\beta}^{FT,r}(x_{1},\ldots,x_{n})=\frac{N!}{(N-n)!}\int_{\Omega_{N-n}}
P_{N\beta}^{FT,r}(x_{1},\ldots,x_{N})d\,\sigma_{N-n}
\end{equation}
where $\Omega_{N-n}$ denotes the sphere
$\sum_{j=n+1}^{N}x^{2}_{j}=r^{2}-\sum_{j=1}^{n}x^{2}_{j}$, and
$d\,\sigma_{N-n}$ denotes the spherical measure. Similarly,  the
$n$-point correlation function for the bounded trace Gaussian
ensembles is defined
 by
\begin{equation}
\label{correlation function for bounded trace}
R_{n\beta}^{BT,r}(x_{1},\ldots,x_{n})=
\frac{N!}{(N-n)!}\int_{\mathbb{R}^{N-n}}
P_{N\beta}^{BT,r}(x_{1},\ldots,x_{N})dx_{n+1}\cdots dx_{N}.
\end{equation}
 Since
\cite{mehta} the expectation of $\sum_{j=1}^{N}x^{2}_{j}$ is
$\frac{N}{4}+(\frac{1}{2 \beta}-\frac{1}{4}) \approx N/4 $ as $N
\rightarrow \infty$, we choose the strength $r=\sqrt{N}/2$ for
restricted trace ensembles and  abbreviate $R_{n\beta}^{FT,r}$ and
$R_{n\beta}^{BT,r}$ to $R_{n\beta}^{FT}$ and $R_{n\beta}^{BT}$. It
is easy to see that the following relation :
\begin{equation}
\label{relation}
r^{n}R_{n\beta}^{FT,r}(rx_{1},\cdots,rx_{n})=R_{n\beta}^{FT,1}(x_{1},\cdots,x_{n}).
\end{equation}

The important thing to be noted about fixed trace GOE, GUE and GSE
is their moment equivalence with the associated Gaussian ensembles
of large dimensions (implying the semicircle law), see Mehta's book
\cite{mehta}, \textbf{Sect}.27.1, p.488. At the end of this section,
p.490, he writes:
\begin{quote}
It is not very clear whether this moment equivalence implies that
all local statistical properties of the eigenvalues in two sets of
ensembles are identical. This is so because these local properties
of eigenvalues may not be expressible only in terms of finite
moments of the matrix elements. \end{quote} In addition, in
\textbf{Sect}.27.3 Mehta also
 speculated that working out
the eigenvalues spacing distribution for bounded trace ensembles
(the semicircle law is also applicable) is much more difficult.
However, to our knowledge, only very few results are obtained on the
local limit behavior of the correlation functions for fixed and
bounded trace ensembles. An important breakthrough about local
scaling limits of the correlation functions for fixed trace
ensembles comes from \cite{av,ggl,gg}. In  \cite{av}, universality
at zero was proved for a class of fixed trace unitary ensembles with
monomial potentials (including Gaussian case). In \cite{ggl},
universality at zero is shown for the correlation measure for fixed
trace GUE and then extended to the bulk \cite{gg} for the
correlation functions using a different method from that in
\cite{ggl}. Recently, the authors have obtained local result at the
edge of the density for Gaussian ensembles in \cite{zlq} and all
kinds of universal properties of correlation functions for fixed and
bounded trace Laguerre unitary ensemble in \cite{lz}. For the
convenience, we state explicitly  universality of sine-kernel in the
bulk for fixed trace GUE as follows:
 for every $u \in (-1,1)$ and
$t_{1},\ldots,t_{n} \in \mathbb{R}^{1}$,
\begin{equation}
\lim_{N \rightarrow \infty}\label{sinekernelforFTGUE}
\frac{1}{(N\omega(u))^{n}}
R^{FT}_{n2}(u+\frac{t_{1}}{N\omega(u)},\ldots,u+\frac{t_{n}}{N\omega(u)})=
\textrm{det}[K_{\textrm{sine}}(t_{j},t_{k})]_{j,k=1}^{n}.
\end{equation}
In the present paper, universality of sine-kernel at zero (in the
bulk for bounded trace GUE) and Airy-kernel at the edge for fixed
and bounded trace GOE, GUE and GSE
 is proved. All these known results, to some extent, answer the problem:
``equivalence of ensembles'' posed by Mehta. However, we must
emphasize that some (macroscopic) universal property from the
unconstrained ensembles might be destroyed when referring to
restricted ensembles, for example, G. Akemann et al \cite {acmv2}
have discovered non-universality of $n$-point resolvent ($n\geq 2$)
for restricted ensembles. %The cluster functions remain universal for the classical ensembles
%GOE, GUE and GSE but not for the corresponding FTE, see discussion e.g. in \cite{av}.

Let $C_{c}(\mathbb{R}^{n})$ be the set of all continuous functions
on $\mathbb{R}^{n}$  with compact support, now we can state our main
results as follows.

\begin{theorem}\label{theoremforFT}
Let $R_{n\beta}^{FT}$ ($\beta=1,2,4$) be the $n$-point correlation
functions of fixed trace GOE, GUE and GSE, defined by
(\ref{correlation function for fixed trace}). For any  $f \in
C_{c}(\mathbb{R}^{n})$, the  following asymptotic properties hold.

(i)At zero of the spectrum:\begin{align}\label{theoremforFTatzero}
&\lim_{N\rightarrow\infty}\int_{\mathbb{R}^{n}}
f(t_{1},\cdots,t_{n})\,(\frac{\pi}{2N})^{n}R_{n\beta}^{FT}(\frac{\pi
t_{1}}{2N},\cdots,\frac{\pi t_{n}}{2N})\,d^{n}t
\nonumber\\
&=\begin{cases}\int_{\mathbb{R}^{n}}
f(t_{1},\cdots,t_{n})\det[K_{\rm{sine}}(t_{j},t_{k})]_{j,k=1}^{n}\,d^{n}t, \ \ \ \ \ \beta=2,\\
\int_{\mathbb{R}^{n}}
f(t_{1},\cdots,t_{n})\big(\det[K_{\rm{sine}}^{(\beta)}(t_{j},t_{k})]_{j,k=1}^{n}\big)^{\frac{1}{2}}
\,d^{n}t, \ \beta=1,4.
\end{cases}
\end{align}

(ii) The soft edge of the spectrum:
\begin{align}\label{theoremforFTatedge}
&\lim_{N\rightarrow\infty}\int_{\mathbb{R}^{n}}
f(t_{1},\cdots,t_{n})\,\frac{1}{(2N^{2/3})^{n}}R_{n\beta}^{FT}(1+\frac{t_{1}}{2N^{2/3}},\cdots,1+\frac{t_{n}}{2N^{2/3}})\,d^{n}t
\nonumber\\
&=\begin{cases}\int_{\mathbb{R}^{n}}
f(t_{1},\cdots,t_{n})\det[K_{\rm{Airy}}(t_{j},t_{k})]_{j,k=1}^{n}\,d^{n}t, \ \ \ \ \ \beta=2,\\
\int_{\mathbb{R}^{n}}
f(t_{1},\cdots,t_{n})\big(\det[K_{\rm{Airy}}^{(\beta)}(t_{j},t_{k})]_{j,k=1}^{n}\big)^{\frac{1}{2}}
\,d^{n}t, \ \beta=1,4.
\end{cases}
\end{align}
\end{theorem}

\begin{theorem}\label{theoremforBT}
Let $R_{n\beta}^{BT}$ ($\beta=1,2,4$) be the $n$-point correlation
functions of bounded trace GOE, GUE and GSE, defined by
(\ref{correlation function for bounded trace}). For any  $f \in
C_{c}(\mathbb{R}^{n})$, the  following asymptotic properties hold.

(i)At zero of the spectrum:\begin{align}\label{theoremforBTatzero}
&\lim_{N\rightarrow\infty}\int_{\mathbb{R}^{n}}
f(t_{1},\cdots,t_{n})\,(\frac{\pi}{2N})^{n}R_{n\beta}^{BT}(\frac{\pi
t_{1}}{2N},\cdots,\frac{\pi t_{n}}{2N})\,d^{n}t
\nonumber\\
&=\begin{cases}\int_{\mathbb{R}^{n}}
f(t_{1},\cdots,t_{n})\det[K_{\rm{sine}}(t_{j},t_{k})]_{j,k=1}^{n}\,d^{n}t, \ \ \ \ \ \beta=2,\\
\int_{\mathbb{R}^{n}}
f(t_{1},\cdots,t_{n})\big(\det[K_{\rm{sine}}^{(\beta)}(t_{j},t_{k})]_{j,k=1}^{n}\big)^{\frac{1}{2}}
\,d^{n}t, \ \beta=1,4.
\end{cases}
\end{align}

(ii) The soft edge of the spectrum:
\begin{align}\label{theoremforBTatedge}
&\lim_{N\rightarrow\infty}\int_{\mathbb{R}^{n}}
f(t_{1},\cdots,t_{n})\,\frac{1}{(2N^{2/3})^{n}}R_{n\beta}^{FT}(1+\frac{t_{1}}{2N^{2/3}},\cdots,1+\frac{t_{n}}{2N^{2/3}})\,d^{n}t
\nonumber\\
&=\begin{cases}\int_{\mathbb{R}^{n}}
f(t_{1},\cdots,t_{n})\det[K_{\rm{Airy}}(t_{j},t_{k})]_{j,k=1}^{n}\,d^{n}t, \ \ \ \ \ \beta=2,\\
\int_{\mathbb{R}^{n}}
f(t_{1},\cdots,t_{n})\big(\det[K_{\rm{Airy}}^{(\beta)}(t_{j},t_{k})]_{j,k=1}^{n}\big)^{\frac{1}{2}}
\,d^{n}t, \ \beta=1,4.
\end{cases}
\end{align}

(iii) The bulk of the spectrum for the bounded trace GUE: for every
$u \in (-1,1)$ and $t_{1},\ldots,t_{n} \in \mathbb{R}^{1}$,
\begin{align}
\lim_{N \rightarrow \infty}&\label{sinekernelforBTGUE}
\int_{\mathbb{R}^{n}}
f(t_{1},\cdots,t_{n})\,\frac{1}{(N\omega(u))^{n}}
R^{BT}_{n2}(u+\frac{t_{1}}{N\omega(u)},\ldots,u+\frac{t_{n}}{N\omega(u)})\,d^{n}t\nonumber\\
&= \int_{\mathbb{R}^{n}}
f(t_{1},\cdots,t_{n})\det[K_{\rm{sine}}(t_{j},t_{k})]_{j,k=1}^{n}\,d^{n}t.
\end{align}
\end{theorem}

The rest of this paper is organized as follows: in Sect. 2, a
relation of correlation functions between fixed trace  and
unconstrained ensembles  is formulated, which is the starting point
of our arguments. Then a useful lemma (see Lemma \ref{lemma1}) is
given, which plays a vital role on the proof of Theorem
\ref{theoremforFT}. Theorems \ref{theoremforFT} and
\ref{theoremforBT} will be proved in Sects. 3 and 4, respectively.

$Notation$. We will use the notation $\vec{v_{n}}$ for the
$n$-dimensional row vector $(v_{1},\ldots,v_{n})$ without further
explanation. The abbreviation $d\vec{v_{n}}$ denotes the Lebesgue
measure $dv_{1}\ldots dv_{n}$ on $\mathbb{R}^{n}$. The symbol
$\|v_{n}\|$ is expressed as the $l^{2}$ (Euclidean) norm of the
vector $\vec{v_{n}}$. We will use the convention that the $C$'s
denote generically bounded constants independent on $N$, whose
values may depend on $\beta$ and change from line to line.

\section{Relation between fixed trace  and unconstrained ensembles}
\label{priminary}

For the $n$-point correlation function of the Gaussian ensembles,
defined by (\ref{correlation function}), and that of the fixed trace
Gaussian ensembles defined by (\ref{correlation function for fixed
trace}), references \cite{acmv,dl,ld} give an integral equation
\begin{equation}
\label{integralequation} R_{n\beta}(x_{1},\ldots,x_{n})=
\frac{1}{C_{N\beta}}(\frac{N}{\sqrt{2}})^{N_{\beta}}\int_{\frac{2}{\sqrt{N}}\|\vec{x_{n}}\|}^{+\infty}e^{-\frac{\beta
u^{2}N^{2}}{4}}\,u^{N_{\beta}-n-1}R_{n\beta}^{FT}(\frac{x_{1}}{u},\ldots,\frac{x_{n}}{u})\,du,
\end{equation}
where
\begin{equation}
\label{changshu}
C_{N\beta}=\Gamma(\frac{N_{\beta}}{2})2^{\frac{N_{\beta}}{2}-1}\beta^{-\frac{N_{\beta}}{2}}.
\end{equation}
Now we give a new argument by a direct calculation. $\forall\, f\in
C_{c}(\mathbb{R}^{n})$,  a transformation to polar coordinates
yields
\begin{align}
&\int_{\mathbb{R}^{n}}f(x_{1},\ldots,x_{n})R_{n\beta}(x_{1},\ldots,x_{n})\,dx_{1}\ldots
dx_{n}\nonumber\\
&=\frac{1}{Z_{N\beta}}\int_{\mathbb{R}^{N}}f(x_{1},\ldots,x_{n})\prod_{1\leq
j<k\leq N}|x_{j}-x_{k}|^{\beta}\prod_{i=1}^{N}e^{-\beta N
x_{i}^{2}}dx_{1}\ldots dx_{N}\nonumber\\
&=\frac{1}{Z_{N\beta}}\int_{0}^{+\infty}\int_{S^{N-1}}f(uy_{1},\ldots,uy_{n})\prod_{1\leq
j<k\leq N}|y_{j}-y_{k}|^{\beta}e^{-\beta N
u^{2}}\,u^{N_{\beta}-1}\,dy_{1}\ldots
dy_{N}\,du\nonumber\\
&=\frac{Z_{N\beta}^{FT,1}}{Z_{N\beta}}\int_{0}^{\infty}\int_{\|\vec{y_{n}}\|\leq
1}f(uy_{1},\cdots,uy_{n})R_{n\beta}^{FT,1}(y_{1},\cdots,y_{n})e^{-\beta
N
u^{2}}\,u^{N_{\beta}-1}\,dy_{1}\ldots dy_{n}\,du\nonumber\\
&=\frac{Z_{N\beta}^{FT,1}}{Z_{N\beta}}\int_{\mathbb{R}^{n}}\int_{\|\vec{x_{n}}\|}^{\infty}f(x_{1},\cdots,x_{n})
R_{n\beta}^{FT,1}(\frac{x_{1}}{u},\ldots,\frac{x_{n}}{u})e^{-\beta N
u^{2}}\,u^{N_{\beta}-n-1}\,du\,dx_{1}\ldots dx_{n}
\end{align}
Combining (\ref{partition function}), (\ref{partition function for
fixed trace}) and (\ref{relation}), we thus conclude
Eq.(\ref{integralequation}).

The integral equation (\ref{integralequation}) as a bridge between
the two ensembles, is the starting point of our arguments. %however
%the local results of correlation functions of fixed trace G$\beta$E
%can't be directly derived from the corresponding G$\beta$E. We will
%give a coarse description about how to obtain the local results of
%fixed trace G$\beta$E based on (\ref{integralequation}) after a
Now we  state a lemma which plays a vital role in the proof of
Theorem \ref{theoremforFT}.
\begin{lemma}
\label{lemma1} Let $\{\alpha_{_{N}}\}$ be a sequence such that
$\alpha_{_{N}}\rightarrow 0$ but $\alpha_{_{N}} N/\sqrt{\ln
N}\rightarrow\infty$ as $N\rightarrow\infty$. Then for any given
$\beta>0$ and nonnegative integer $n$, as $N\rightarrow\infty$, we
have
\begin{align}
\frac{1}{C_{N\beta}}\int_{0}^{\frac{N}{\sqrt{2}}(1-\alpha_{_{N}})}e^{-\frac{\beta\,u^{2}}{2}}u^{N_{\beta}-n-1}\,du=O(e^{-0.5\beta
(N\alpha_{_{N}})^{2}(1+o(1))}) \label{estimate2}
\end{align}
and
\begin{align}
\frac{1}{C_{N\beta}}\int_{\frac{N}{\sqrt{2}}(1+\alpha_{_{N}})}^{+\infty}e^{-\frac{\beta
u^{2}}{2}}u^{N_{\beta}-n-1}\,du=O(e^{-0.5\beta
(N\alpha_{_{N}})^{2}(1+o(1))})\label{estimate3}
\end{align}
where $C_{N\beta}$ is given by (\ref{changshu}).
\end{lemma}
\begin{proof}We will first consider the left hand side of (\ref{estimate2}). It is not
difficult to observe that the function $e^{-\beta
u^{2}/2}u^{N_{\beta}-n-1}$ attains its maximum at
$\sqrt{(N_{\beta}-n-1)/\beta}$ which satisfies
\begin{equation}
\label{part1maximunpoint}
\sqrt{\frac{N_{\beta}-n-1}{\beta}}>\frac{N}{\sqrt{2}}(1-\alpha_{_{N}})
\end{equation}
for sufficiently large $N$. Therefore,  using
 Stirling's formula  for the gamma function:
\begin{equation}
\Gamma(\frac{N_{\beta}}{2})
=(2\pi)^{1/2}e^{-N_{\beta}/2}\,(\frac{N_{\beta}}{2})^{\frac{N_{\beta}-1}{2}}(1+O(\frac{1}{N})),
\end{equation}
one obtains
\begin{align}
\label{d1}
\frac{1}{C_{N\beta}}\int_{0}^{\frac{N}{\sqrt{2}}(1-\alpha_{_{N}})}e^{-\frac{\beta\,u^{2}}{2}}u^{N_{\beta}-n-1}\,du
\leq \frac{e^{-\frac{\beta
N^{2}}{4}(1-\alpha_{_{N}})^{2}}(\frac{N}{\sqrt{2}}(1-\alpha_{_{N}}))^{N_{\beta}-n-1}}
{\Gamma(N_{\beta}/2)2^{N_{\beta}/2-1}\beta^{-N_{\beta}/2}}\,\frac{N}{\sqrt{2}}(1-\alpha_{_{N}})\nonumber\\
\leq C\,\frac{e^{-\frac{\beta
N^{2}}{4}(1-\alpha_{_{N}})^{2}}2^{-\frac{N_{\beta}-n-1}{2}}(1-\alpha_{_{N}})^{N_{\beta}-n-1}}{e^{-N_{\beta}/2}}\,
\frac{(\sqrt{\beta}N)^{N_{\beta}}}{N_{\beta}^{N_{\beta}/2}}\,N^{1-n}.
\end{align}
Expanding
$$\ln (
N_{\beta}^{N_{\beta}/2})=\frac{N_{\beta}}{2}\big(\ln(\frac{\beta}{2}N^{2})+\ln(1+(\frac{2}{\beta}-1)
\frac{1}{N})\big)$$ and by a direct calculation, one obtains
\begin{equation}
\frac{(\sqrt{\beta}N)^{N_{\beta}}}{N_{\beta}^{N_{\beta}/2}}=\exp(\frac{N_{\beta}}{2}\ln
2-\frac{N}{2}+\frac{\beta N}{4}+O(1)).\nonumber
\end{equation}
Again, expanding
$$\ln (1-\alpha_{_{N}})=-\alpha_{_{N}}-\alpha_{_{N}}^{2}/2+O(\alpha_{_{N}}^{3}),$$
thus (\ref{d1}) can be dominated by
\begin{align}
\label{d1result} &(\ref{d1})\leq C\exp\Big(-\frac{\beta
N^{2}}{4}(1-\alpha_{_{N}})^{2}-\frac{N_{\beta}-n-1}{2}\ln2+(N_{\beta}-n-1)\ln(1-\alpha_{_{N}})
+\frac{N_{\beta}}{2}\nonumber\\&+\frac{N_{\beta}}{2}\ln2-\frac{N}{2}+
\frac{\beta N}{4}+(1-n)\ln N\Big)\nonumber\\
=&C\exp\big(-0.5\beta
(N\alpha_{_{N}})^{2}+\frac{n+1}{2}\ln2+(1-n)\ln
N+(N(1-\frac{\beta}{2})-n-1)\ln(1-\alpha_{_{N}})+O(\alpha_{_{N}}^{3}N^{2})\big)\nonumber\\
= &C\exp\big(-0.5\beta
(N\alpha_{_{N}})^2(1+O(\alpha_{_{N}})+O((\alpha_{_{N}}\,N)^{-1}))+
O((\alpha_{_{N}}\,N)^{-2}))+O((\alpha_{_{N}}N)^{-2}\ln N) \big)\nonumber\\
=&C\exp\big(-0.5\beta (N\alpha_{_{N}})^2(1+o(1))\big).
\end{align}
Here we make use of the assumption about the sequence
$\alpha_{_{N}}$. Following the similar argument, (\ref{estimate3})
can be obtained. Indeed, for sufficiently large $N$, the function
$e^{-u^{2}\beta/2}u^{N_{\beta}+1}$ attains its maximum at
$\sqrt{(N_{\beta}+1)/\beta}$ with the condition
\begin{equation}
\sqrt{(N_{\beta}+1)/\beta}<\frac{N}{\sqrt{2}}(1+\alpha_{_{N}}).\nonumber
\end{equation}
Thus we have
\begin{align}
\label{d2result}
&\frac{1}{C_{N\beta}}\int_{\frac{N}{\sqrt{2}}(1+\alpha_{_{N}})}^{\infty}e^{-\frac{\beta
u^{2}}{2}}u^{N_{\beta}+1}u^{-n-2}du \leq\,\frac{e^{-\frac{\beta
N^{2}}{4}(1+\alpha_{_{N}})^{2}}(\frac{N}{\sqrt{2}}(1+\alpha_{_{N}}))^{N_{\beta}+1}}{\Gamma(N_{\beta}/2)2^{N_{\beta}/2-1}\beta^{-N_{\beta}/2}}\,
\int_{\frac{N}{\sqrt{2}}(1+\alpha_{_{N}})}^{\infty}u^{-n-2}du\nonumber\\
&\leq C\exp\big(-0.5\beta
(N\alpha_{_{N}})^{2}-\frac{1}{2}\ln2+(1-n)\ln
N+(N(1-\frac{\beta}{2})+1)\ln(1+\alpha_{_{N}})+O(\alpha_{_{N}}^{2}N^{3})
\big)\nonumber\\ =&C\exp\big(-0.5\beta
(N\alpha_{_{N}})^2(1+o(1))\big).
\end{align}
This completes the proof of this lemma.
\end{proof}
\begin{remark}
For the convenience of our arguments, set
\begin{equation}
\label{abbrivation}
\Psi_{N\beta}(u)\triangleq\frac{1}{C_{N\beta}}(\frac{N}{\sqrt{2}})^{N_{\beta}}\,e^{-\frac{\beta
u^{2}N^{2}}{4}}\,u^{N_{\beta}-1}.
\end{equation}
Thus
\begin{align}
\int_{0}^{1-\alpha_{_{N}}}\Psi_{N\beta}(u)du=\frac{1}{C_{N\beta}}\int_{0}^{\frac{N}{\sqrt{2}}(1-\alpha_{_{N}})}e^{-\frac{\beta\,u^{2}}{2}}u^{N_{\beta}-1}\,du=O(e^{-0.5\beta
(N\alpha_{_{N}})^{2}(1+o(1))}), \label{abbribation 1}\\
\int_{1+\alpha_{_{N}}}^{\infty}\Psi_{N\beta}(u)du=\frac{1}{C_{N\beta}}\int_{\frac{N}{\sqrt{2}}(1+\alpha_{_{N}})}^{+\infty}e^{-\frac{\beta
u^{2}}{2}}u^{N_{\beta}-1}\,du=O(e^{-0.5\beta
(N\alpha_{_{N}})^{2}(1+o(1))}). \label{abbribation 2}
\end{align}
Here we make use of the above lemma in the case: $n=0$. It is very
easy to see that
\begin{align}
\label{intermediateintegral}
\int_{1-\alpha_{_{N}}}^{1+\alpha_{_{N}}}\Psi_{N\beta}(u)du=1+O(e^{-0.5\beta
(N\alpha_{_{N}})^{2}(1+o(1))}).
\end{align}
\end{remark}
By (\ref{abbrivation}), the integral equation
(\ref{integralequation}) can be rewritten as follows:
\begin{equation}
\label{integral equation: abbrevation} R_{n\beta}(\vec{x_{n}})=
\int_{\frac{2}{\sqrt{N}}\|\vec{x_{n}}\|}^{+\infty}\Psi_{N\beta}(u)\frac{1}{u^{n}}R_{n\beta}^{FT}(\frac{1}{u}\vec{x_{n}})\,du.
\end{equation}

At the end of this section, we give a simple sketch of the proof of
Theorem \ref{theoremforFT}. Notice the fact that the right hand side
of the above equation can be divided into three parts, i.e.,
\begin{equation}
\label{3.12 equals}
\left(\int_{\frac{2}{\sqrt{N}}\|\vec{x_{n}}\|}^{1-\alpha_{_{N}}}+\int_{1-\alpha_{_{N}}}^{1+\alpha_{_{N}}}
+\int_{1+\alpha_{_{N}}}^{\infty}\right)
\Psi_{N\beta}(u)\frac{1}{u^{n}}R_{n\beta}^{FT}(\frac{1}{u}\vec{x_{n}})\,du.
\end{equation}
Using Lemma \ref{lemma1}, the first and third parts  will rapidly
 disappear for the large $N$. By the integral intermediate value theorem, the middle part of the above identity equals
\begin{equation}\int_{1-\alpha_{_{N}}}^{1+\alpha_{_{N}}}\Psi_{N\beta}(u)\,du\
\frac{1}{\big(\xi_{N}(u)\big)^{n}}R_{n\beta}^{FT}(\frac{1}{\xi_{N}(u)}\vec{x_{n}})\end{equation}
where $1-\alpha_{_{N}}\leq \xi_{N}(u)\leq 1+\alpha_{_{N}}$. It
follows from $1/\xi_{N}(u)=1+O(\alpha_{_{N}})$ that
\begin{equation}R_{n \beta}(\vec{x_{n}})\sim
R_{n\beta}^{FT}((1+O(\alpha_{_{N}}))\vec{x_{n}}).\end{equation}
 Let $x_{i}=\pi
t_{i}/2N,\, 1\leq i\leq n$, then the above relation can be rewritten
as \begin{equation}R_{n\beta}(\pi t_{1}/2N,\ldots,\pi t_{1}/2N)\sim
R_{n\beta}^{FT}\Big((1+O(\alpha_{_{N}}))\pi
t_{1}/2N,\ldots,(1+O(\alpha_{_{N}}))\pi t_{1}/2N\Big).\end{equation}
It is not difficult to see that $(1+O(\alpha_{_{N}}))\pi
t_{1}/2N=\pi t_{1}/2N+O(\alpha_{_{N}}/N).$ The fact that
$\alpha_{_{N}}\rightarrow 0$ as $N$ goes to infinity means that
$\alpha_{_{N}}/N=o(1/N)$. Therefore the term $O(\alpha_{_{N}}/N)$ is
a small perturbation, comparing with $\pi t_{1}/2N$, and the
relation \begin{equation}R_{n\beta}(\pi t_{1}/2N,\ldots,\pi
t_{1}/2N)\sim R_{n\beta}^{FT}(\pi t_{1}/2N,\ldots,\pi
t_{1}/2N)\end{equation} can be expected. Similarly, at the edge of
the spectrum, we assume that $x_{i}=1+ t_{i}/2N^{2/3},\, 1\leq i\leq
n$, then
\begin{equation}R_{n\beta}(1+ t_{1}/2N^{2/3},\ldots,1+ t_{i}/2N^{2/3})\sim
R_{n\beta}^{FT}\Big((1+O(\alpha_{_{N}}))(1+
t_{1}/2N^{2/3}),\ldots,(1+O(\alpha_{_{N}}))(1+
t_{1}/2N^{2/3})\Big).\end{equation} Notice
$$(1+O(\alpha_{_{N}}))(1+ t_{i}/2N^{2/3})=1+ t_{i}/2N^{2/3}+ O(\alpha_{_{N}})(1+ t_{i}/2N^{2/3})
,$$ and if one takes $\alpha_{_{N}}=N^{-\theta},\ 2/3<\theta<1$,
then the relation
\begin{equation}R_{n\beta}(1+ t_{1}/2N^{2/3},\ldots,1+ t_{i}/2N^{2/3})\sim
R_{n\beta}^{FT}\Big(1+ t_{1}/2N^{2/3},\ldots,1+
t_{i}/2N^{2/3}\Big)\end{equation} can also be expected.

 %Thus if  the middle part could be
% rewritten approximately as
% $$\int_{1-\alpha_{_{N}}}^{1+\alpha_{_{N}}}\Psi_{N\beta}(u)\,du\ \frac{1}{u^{n}}R_{n\beta}^{FT}(\frac{1}{u}\vec{x_{n}}),$$
% then by (\ref{intermediateintegral}) and the range of $u$ the proofs can be completed.
 In the present paper, we mainly use the case
$\alpha_{_{N}}=N^{-\theta}$ for any fixed  $\frac{2}{3}<\theta<1$,
which obviously satisfies the assumption of Lemma \ref{lemma1} about
$\alpha_{_{N}}$. %Now we turn to the proof of Theorem $\ref{global limit

\section{Proof of theorem \ref{theoremforFT} }
 In principle, universality of sine-kernel at zero can be
proved by one of the arguments \cite{zlq}: giving  an upper bound
estimation of the correlation functions with the help of the maximum
of Vandermonde determinant on the sphere by Stieltjes \cite{szego}.
However, in this paper we will deal with it in a slightly different
way. %But our arguments seem to be insufficient for proving
%universality in the whole bulk.
From Lemma \ref{lemma1}, let us take $\alpha_{_{N}}=N^{-\theta},
\theta\in (0,1)$. It is a well-known fact that the scaling at the
soft edge of the spectrum is proportional to $N^{-2/3}$, thus we can
choose $\theta>2/3$ and give a very close approximation of
correlation functions near the radial sharp cutoff point. Then using
known results about the unconstrained ensembles, we can obtain
Airy-kernel for the fixed trace ensembles. However, this argument
seems to be insufficient for proving universality in the whole bulk.
The main difficulty is that the ``rate" index $\theta$ has been
rather sharp, in the sense that it cannot be replaced with a larger
number than 1.

%Theorem 2 is given,
\begin{proof}[Proof of Theorem \ref{theoremforFT}:(i)]
By (\ref{integral equation: abbrevation}), we obtain
\begin{align}
\label{divideintegral:at zero}&(\frac{\pi}{2N})^{n}
\int_{\mathbb{R}^{n}}f(\vec{t_{n}})R_{n\beta}(\frac{\pi}{2N}\vec{t_{n}})d\,\vec{t_{n}}
=(\frac{\pi}{2N})^{n}\int_{\mathbb{R}^{n}}
f(\vec{t_{n}})\int_{\frac{\pi}{\sqrt{N^{3}}}\|\vec{t_{n}}\|}^{+\infty}
\Psi_{N\beta}(u)\,\frac{1}{u^{n}}\,R_{n\beta}^{FT}(\frac{\pi}{2Nu}\vec{t_{n}})du\,\vec{t_{n}}\nonumber\\
&=(\frac{\pi}{2N})^{n}\int_{\mathbb{R}^{n}} f(\vec{t_{n}})
\Big(\int_{\frac{\pi}{\sqrt{N^{3}}}\|\vec{t_{n}}\|}^{1-\alpha_{_{N}}}+\int_{1-\alpha_{_{N}}}^{1+\alpha_{_{N}}}+\int_{1+\alpha_{_{N}}}^{\infty}\Big)
\Psi_{N\beta}(u)\,\frac{1}{u^{n}}\,R_{n\beta}^{FT}(\frac{\pi}{2Nu}\vec{t_{n}})du\,\vec{t_{n}}\nonumber\\
&=I+II+III.
\end{align}
%where
%\begin{align}
%&I=(\frac{\pi}{2N})^{n}\int_{\mathbb{R}^{n}}
%f(\vec{t_{n}})\int_{\frac{\pi}{\sqrt{N^{3}}}\|\vec{t_{n}}\|}^{1-\alpha_{_{N}}}
%\Psi_{N\beta}(u)\,\frac{1}{u^{n}}\,R_{n\beta}^{FT}(\frac{\pi}{2Nu}\vec{t_{n}})du\,d\vec{t_{n}}\\
%&II=(\frac{\pi}{2N})^{n}\int_{\mathbb{R}^{n}}
%f(\vec{t_{n}})\int_{1-\alpha_{_{N}}}^{1+\alpha_{_{N}}}
%\Psi_{N\beta}(u)\,\frac{1}{u^{n}}\,R_{n\beta}^{FT}(\frac{\pi}{2Nu}\vec{t_{n}})du\,d\vec{t_{n}}\\
%&III=(\frac{\pi}{2N})^{n}\int_{\mathbb{R}^{n}}
%f(\vec{t_{n}})\int_{1+\alpha_{_{N}}}^{\infty}
%\Psi_{N\beta}(u)\,\frac{1}{u^{n}}\,R_{n\beta}^{FT}(\frac{\pi}{2Nu}\vec{t_{n}})du\,d\vec{t_{n}}
%\end{align}
We will first estimate $I$. Making the change of variables
\begin{equation}
\label{variable subs:at zero} t_{i}=u\,y_{i},\,1\leq i\leq n,
\end{equation} then $I$ can be reduced to
\begin{align}
(\frac{\pi}{2N})^{n}\int_{\mathbb{R}^{n}}\int_{0}^{1-\alpha_{_{N}}}
f(u\vec{y_{n}})1_{\{\frac{\pi}{\sqrt{N^{3}}}\|\vec{y_{n}}\|\leq
u\leq
1-\alpha_{_{N}}\}}(u)\,\Psi_{N\beta}(u)R_{n\beta}^{FT}(\frac{\pi
\vec{y_{n}}}{2N})du\,d\vec{y_{n}}.\end{align} Since $f\in
C_{c}(\mathbb{R}^{n})$, there exists some positive constant $M$ such
that $|f(x)|\leq M$. It follows from (\ref{abbribation 1}) that $I$
can be dominated by
\begin{align}
\label{I} I&\leq
M(\frac{\pi}{2N})^{n}\int_{\mathbb{R}^{n}}\int_{0}^{1-\alpha_{_{N}}}
\Psi_{N\beta}(u)\,R_{n\beta}^{FT}(\frac{\pi
\vec{y_{n}}}{2N})du\,d\vec{y_{n}}\nonumber\\
&=M\int_{0}^{1-\alpha_{_{N}}}\Psi_{N\beta}(u)du\,
(\frac{\pi}{2N})^{n}\int_{\mathbb{R}^{n}} R_{n\beta}^{FT}(\frac{\pi
\vec{y_{n}}}{2N})d\vec{y_{n}}=M\int_{0}^{1-\alpha_{_{N}}}\Psi_{N\beta}(u)du\int_{\mathbb{R}^{n}}
R_{n\beta}^{FT}(\vec{z_{n}})d\vec{z_{n}}\nonumber\\
&=O(N^{n}\,e^{-0.5\beta N^{2(1-\theta)}(1+o(1))}).
\end{align}
Here we make the change of variables: $z_{i}=(2N)/\pi y_{i},\, 1\leq
i\leq n$ for the integral
$$(\frac{\pi}{2N})^{n}\int_{\mathbb{R}^{n}}
R_{n\beta}^{FT}(\frac{\pi \vec{y_{n}}}{2N})\,d\vec{y_{n}}.$$ On the
other hand, the fact that $\int_{\mathbb{R}^{n}}
R_{n\beta}^{FT}(\vec{z_{n}})d\vec{z_{n}}=N!/(N-n)!$ has been used.
According to (\ref{abbribation 2}), a similar argument shows that
$III$ can be dominated by
\begin{align}
\label{III} III&\leq
M(\frac{\pi}{2N})^{n}\int_{\mathbb{R}^{n}}\int_{1+\alpha_{_{N}}}^{\infty}
\Psi_{N\beta}(u)R_{n\beta}^{FT}(\frac{\pi
\vec{y_{n}}}{2N})dud\vec{y_{n}}=O(N^{n}\,e^{-0.5\beta
N^{2(1-\theta)}(1+o(1))}).
\end{align}
Next, we will consider $II$. Under the transform (\ref{variable
subs:at zero}), we find
\begin{align}
\label{theorem2:partII} II=(\frac{\pi}{2N})^{n}\int_{\mathbb{R}^{n}}
\int_{1-\alpha_{_{N}}}^{1+\alpha_{_{N}}}
f(u\,\vec{y_{n}})\Psi_{N\beta}(u)R_{n\beta}^{FT}(\frac{\pi}{2N}\vec{y_{n}})dud\vec{y_{n}}.\end{align}
Combining (\ref{divideintegral:at zero}), (\ref{I}) and (\ref{III}),
it is not difficult to see that that for any fixed $f(x)\in
C_{c}(\mathbb{R})$,
\begin{align}
\label{relation: between II and GbetaE bulk at zero }
&(\frac{\pi}{2N})^{n}\int_{\mathbb{R}^{n}}f(\vec{t_{n}})R_{n\beta}(\frac{\pi}{2N}\vec{t_{n}})d\vec{t_{n}}\nonumber\\
&=(\frac{\pi}{2N})^{n}\int_{\mathbb{R}^{n}}
\int_{1-\alpha_{_{N}}}^{1+\alpha_{_{N}}}
f(u\,\vec{y_{n}})\Psi_{N\beta}(u)R_{n\beta}^{FT}(\frac{\pi}{2N}\vec{y_{n}})dud\vec{y_{n}}+O(N^{n}\,e^{-0.5\beta
N^{2(1-\theta)}(1+o(1))}).
\end{align}
Observe that
\begin{align}
\label{result:at zero} &(\frac{\pi}{2N})^{n}\int_{\mathbb{R}^{n}}
\int_{1-\alpha_{_{N}}}^{1+\alpha_{_{N}}}
f(\vec{y_{n}})\Psi_{N\beta}(u)R_{n\beta}^{FT}(\frac{\pi}{2N}\vec{y_{n}})dud\vec{y_{n}}\\
&=(\frac{\pi}{2N})^{n}\int_{\mathbb{R}^{n}}
f(\vec{y_{n}})R_{n\beta}^{FT}(\frac{\pi\vec{y_{n}}}{2N})du\,d\vec{y_{n}}(1+O(N^{n}\,e^{-0.5\beta
N^{2(1-\theta)}(1+o(1))})).\nonumber
\end{align}
Next we will prove that the difference between (\ref{result:at
zero}) and (\ref{theorem2:partII}) is zero as $N$ goes to infinity,
i.e.
\begin{align}
\label{difference:at zero}
\lim_{N\rightarrow\infty}\big|(\frac{\pi}{2N})^{n}\int_{\mathbb{R}^{n}}
\int_{1-\alpha_{_{N}}}^{1+\alpha_{_{N}}}\big(f(u\vec{y_{n}})-f(\vec{y_{n}})\big)\Psi_{N\beta}(u)
R_{n\beta}^{FT}(\frac{\pi\vec{y_{n}}}{2N})du\,d\vec{y_{n}}\big|=0.
\end{align}
Note that $f(\vec{x})\in C_{c}(\mathbb{R}^{n})$. For any
$\epsilon>0$, there exists some $\delta(\epsilon)>0$ such that
\begin{equation} |f(\vec{x})-f(\vec{y})|<\epsilon. \nonumber
\end{equation}
whenever $\|\vec{x}-\vec{y}\|<\delta$. We remind that
$\mathbf{B}_{R}$ denotes the ball of the radius $R$ in
$\mathbb{R}^{n}$ centered at zero. Choose a ball $\mathbf{B}_{R}$
such that $supp(f)\subset\mathbf{B}_{R}$ and
\{$u\vec{y_{n}}|\vec{y_{n}}\in supp(f),1-\alpha_{_{N}}\leq
u\leq1+\alpha_{_{N}}$\}$\subset\mathbf{B}_{R}$. $\forall\
\vec{y_{n}}\in supp(f)$, there exists $N_{0}$ not depending on
$\vec{y_{n}}$ such that
$$\|u\vec{y_{n}}-\vec{y_{n}}\|\leq |u-1|\|\vec{y_{n}}\|<
N^{-\theta}R<\delta(\epsilon)$$ for $N>N_{0}$. Therefore, we have
\begin{equation}
|f(u\vec{y_{n}})-f(\vec{y_{n}})|<\epsilon.
\end{equation}
For large $N$, it follows from (\ref{intermediateintegral}) that
\begin{align}
\label{dominate} &|(\ref{result:at zero})-
(\ref{theorem2:partII})|\leq(\frac{\pi}{2N})^{n}\int_{\mathbf{B}_{R}}
\int_{1-\alpha_{_{N}}}^{1+\alpha_{_{N}}}|f(u\,\vec{y_{n}})-f(\vec{y_{n}})|\Psi_{N\beta}(u)\,R_{n\beta}^{FT}(\frac{\pi\vec{y_{n}}}{2N})dud\vec{y_{n}}\nonumber\\
&=\epsilon\,
(\frac{\pi}{2N})^{n}\int_{\mathbf{B}_{R}}R_{n\beta}^{FT}(\frac{\pi\vec{y_{n}}}{2N})d\vec{y_{n}}
\int_{1-\alpha_{_{N}}}^{1+\alpha_{_{N}}}\Psi_{N\beta}(u)du \leq
2\epsilon\,(\frac{\pi}{2N})^{n}\int_{\mathbf{B}_{R}}R_{n\beta}^{FT}(\frac{\pi\vec{y_{n}}}{2N})d\vec{y_{n}}\nonumber\\
& \leq 2\epsilon\,M_{_{R}}.
\end{align}
Here we have used Lemma \ref{lemma2} below. Thus,
(\ref{difference:at zero}) can be obtained. The proof of  Lemma
\ref{lemma2} will be postponed until  completing the proof.
According to (\ref{relation: between II and GbetaE bulk at zero })
and (\ref{result:at zero}), it follows that
\begin{align}
(\frac{\pi}{2N})^{n}\int_{\mathbb{R}^{n}}f(\vec{t_{n}})R_{n\beta}(\frac{\pi}{2N}\vec{t_{n}})d\vec{t_{n}}
=(\frac{\pi}{2N})^{n}\int
f(\vec{y_{n}})R_{n\beta}^{FT}(\frac{\pi\vec{y_{n}}}{2N})du\,d\vec{y_{n}}(1+o(1))+o(1).
\end{align}
Hence, from the known results (\ref{sinekernel}) for the Gassian
ensembles, we complete the proof of Theorem \ref{theoremforFT}: (i).
\end{proof}

Now we will prove the following lemma, which is inspired by Lemma 4
in \cite{ggl}.
\begin{lemma}
\label{lemma2} Let $\beta=1,2,4$, for any fixed $R>0$, there exists
a constant $M_{_{R}}$ such that
\begin{equation}
\label{uniform bounded: at zero}
(\frac{\pi}{2N})^{n}\int_{\mathbf{B}_{R}}R_{n\beta}^{FT}(\frac{\pi}{2N}y_{1},\ldots,\frac{\pi}{2N}y_{n})dy_{1}\cdots
dy_{n}\leq M_{_{R}}.
\end{equation}

\end{lemma}
\begin{proof}
 Choose  any positive $\delta<R$, there exists
$N_{o}$ depending only on $R$ and $\delta$ such that
$R\,N^{-\theta}<\delta$ for $N>N_{0}$. That is, for any
$\vec{y_{n}}\in\mathbf{B}_{R}$, when
$u\in[1-N^{-\theta},1+N^{-\theta}]$, we have
$$u\|\vec{y_{n}}\|\leq R+N^{-\theta}R < R+\delta.$$
Let $\eta\in(0,1)$ be a real number and let $\phi(t)$ be a smooth
decreasing function on $[0,R+\delta)$ such that $\phi(t) = 1$ for $t
\in[0,R+\delta)$ and $\phi(t)=0$ for $t\geq(1+\eta)(R+\delta)$. Set
$\varphi(\vec{x_{n}})=\phi(\|\vec{x_{n}}\|)$ for $\vec{x_{n}} =
(x_{1},\cdots, x_{n})\in\mathbb{R}^{n}$.  For $N>N_{0}$, we have
\begin{equation}
\label{lemm2:key scaling}
(\frac{\pi}{2N})^{n}\int_{\mathbf{B}_{R}}R_{n\beta}^{FT}(\frac{\pi\vec{y_{n}}}{2N})d\vec{y_{n}}\leq
(\frac{\pi}{2N})^{n}\int_{\mathbb{R}^{n}}\varphi(u\vec{y_{n}})R_{n\beta}^{FT}(\frac{\pi\vec{y_{n}}}{2N})d\vec{y_{n}}.
\end{equation}
Multiplying by $\Psi_{N\beta}(u)$ then integrating (\ref{lemm2:key
scaling}) with respect to $u$ on
$[1-\alpha_{_{N}},1+\alpha_{_{N}}]$, one obtains
\begin{align}
\label{proposition keyinequality} &
\int_{1-\alpha_{_{N}}}^{1+\alpha_{_{N}}}\Psi_{N\beta}(u)du\,(\frac{\pi}{2N})^{n}\int_{\mathbf{B}_{R}}R_{n\beta}^{FT}(\frac{\pi\vec{y_{n}}}{2N})d\vec{y_{n}}
\nonumber\\&\leq (\frac{\pi}{2N})^{n}\int_{\mathbb{R}^{n}}
\int_{1-\alpha_{_{N}}}^{1+\alpha_{_{N}}}
\varphi(u\vec{y_{n}})\Psi_{N\beta}(u)R_{n\beta}^{FT}(\frac{\pi\vec{y_{n}}}{2N})dud\vec{y_{n}}\nonumber\\
&=(\frac{\pi}{2N})^{n}\int_{\mathbb{R}^{n}}
\varphi(\vec{y_{n}})R_{n\beta}(\frac{\pi}{2N}\vec{y_{n}})d\vec{y_{n}}+O(N^{n}\,e^{-0.5\beta
N^{2(1-\theta)}(1+o(1))}).
\end{align}
Here we have used (\ref{relation: between II and GbetaE bulk at zero
}) and (\ref{intermediateintegral}). From (\ref{sinekernel}), we
conclude (\ref{uniform bounded: at zero}).
\end{proof}

Next, we will prove the part (ii) of Theorem \ref{theoremforFT}.
\begin{proof}[Proof of Theorem \ref{theoremforFT}:(ii)]
Set
$\vec{w_{n}}=(1+\frac{t_{1}}{2N^{2/3}},\cdots,1+\frac{t_{n}}{2N^{2/3}})$.
By (\ref{integral equation: abbrevation}), we find
\begin{align}
\label{theorem2:divided integral}
&\frac{1}{(2N^{2/3})^{n}}\int_{\mathbb{R}^{n}}
f(t_{1},\cdots,t_{n})R_{n\beta}(1+\frac{t_{1}}{2N^{2/3}},\cdots,1+\frac{t_{n}}{2N^{2/3}})dt_{1}\cdots
dt_{n}\nonumber\\
&=\frac{1}{(2N^{2/3})^{n}}\int_{\mathbb{R}^{n}}
f(\vec{t_{n}})\int_{\frac{2}{\sqrt{N}}\|\vec{w_{n}}\|}^{+\infty}\Psi_{N\beta}(u)\frac{1}{u^{n}}R_{n\beta}^{FT}(\frac{\vec{w_{n}}}{u})dud\vec{t_{n}}\nonumber\\
&=\frac{1}{(2N^{2/3})^{n}}\int_{\mathbb{R}^{n}} f(\vec{t_{n}})
\Big(\int_{\frac{2}{\sqrt{N}}\|\vec{w_{n}}\|}^{1-\alpha_{_{N}}}+\int_{1-\alpha_{_{N}}}^{1+\alpha_{_{N}}}+\int_{1+\alpha_{_{N}}}^{\infty}\Big)
\Psi_{N\beta}(u)\frac{1}{u^{n}}R_{n\beta}^{FT}(\frac{\vec{w_{n}}}{u})dud\vec{t_{n}}\nonumber\\
&=I'+II'+III'.
\end{align}
%where
%\begin{align}
%&I'=\frac{1}{(2N^{2/3})^{n}}\int_{\mathbb{R}^{n}}
%f(\vec{t_{n}})\int_{\frac{2}{\sqrt{N}}\|\vec{w_{n}}\|}^{1-\alpha_{_{N}}}\Psi_{N\beta}(u)\frac{1}{u^{n}}R_{n\beta}^{FT}(\frac{\vec{w_{n}}}{u})dud\vec{t_{n}}\\
%&II'=\frac{1}{(2N^{2/3})^{n}}\int_{\mathbb{R}^{n}}
%f(\vec{t_{n}})\int_{1-\alpha_{_{N}}}^{1+\alpha_{_{N}}}\Psi_{N\beta}(u)\frac{1}{u^{n}}R_{n\beta}^{FT}(\frac{\vec{w_{n}}}{u})dud\vec{t_{n}}\\
%&III'=\frac{1}{(2N^{2/3})^{n}}\int_{\mathbb{R}^{n}}
%f(\vec{t_{n}})\int_{1+\alpha_{_{N}}}^{+\infty}\Psi_{N\beta}(u)\frac{1}{u^{n}}R_{n\beta}^{FT}(\frac{\vec{w_{n}}}{u})dud\vec{t_{n}}.
%\end{align}
We will follow an analogous  argument for the proof  of Theorem
\ref{theoremforFT}:(i). Making the change of variables
\begin{equation}
\label{variable subs:edge} t_{i}=uy_{i}+2N^{2/3}(u-1), \, 1\leq
i\leq n,
\end{equation}
then $I'$ can be reduced to
\begin{align}
\label{theorem2:partI} (\frac{1}{2N^{2/3}})^{n}\int_{\mathbb{R}^{n}}
\int_{0}^{1-\alpha_{_{N}}}
f\Big(u\,y_{1}+2N^{2/3}(u-1),\ldots,uy_{n}+2N^{2/3}(u-1)\Big)\nonumber\\
\times 1_{\{\frac{2}{\sqrt{N}}\|\vec{y_{n}}\|\leq u\leq
1-\alpha_{_{N}}\}}(u)\, \Psi_{N\beta}(u)
R_{n\beta}^{FT}(1+\frac{y_{1}}{2N^{2/3}},\ldots,1+\frac{y_{n}}{2N^{2/3}})dud\vec{y_{n}}.
\end{align}
Assume that $|f(x)|\leq M$. By (\ref{abbribation 1}), $I'$ can be
dominated by
\begin{align}
\label{theorem2:partI result} I'&\leq
M(\frac{1}{2N^{2/3}})^{n}\int_{\mathbb{R}^{n}}
\int_{0}^{1-\alpha_{_{N}}}\Psi_{N\beta}(u)R_{n\beta}^{FT}(1+\frac{y_{1}}{2N^{2/3}},\ldots,1+\frac{y_{n}}{2N^{2/3}})dud\vec{y_{n}}\nonumber\\
&=
M(\frac{1}{2N^{2/3}})^{n}\int_{\mathbb{R}^{n}}R_{n\beta}^{FT}(1+\frac{y_{1}}{2N^{2/3}},\ldots,1+\frac{y_{n}}{2N^{2/3}})d\vec{y_{n}}
\int_{0}^{1-\alpha_{_{N}}}\Psi_{N\beta}(u)du\nonumber\\
&=M\frac{N!}{(N-n)!}O(e^{-0.5\beta
N^{2(1-\theta)}(1+o(1))})=O(N^{n}\,e^{-0.5\beta
N^{2(1-\theta)}(1+o(1))}).
\end{align}
Here we use the fact that
\begin{equation}
(\frac{1}{2N^{2/3}})^{n}\int_{\mathbb{R}^{n}}R_{n\beta}^{FT}(1+\frac{y_{1}}{2N^{2/3}},\ldots,1+\frac{y_{n}}{2N^{2/3}})d\vec{y_{n}}=\frac{N!}{(N-n)!}.
\end{equation}
Similarly, by (\ref{abbribation 2}), one gets
\begin{align}
\label{theorem2:partIII} III'&\leq
M(\frac{1}{2N^{2/3}})^{n}\int_{\mathbb{R}^{n}}
\int_{1+\alpha_{_{N}}}^{\infty}\Psi_{N\beta}(u)R_{n\beta}^{FT}(1+\frac{y_{1}}{2N^{2/3}},\ldots,1+\frac{y_{n}}{2N^{2/3}})dud\vec{y_{n}}\nonumber\\
&=O(N^{n}\,e^{-0.5\beta N^{2(1-\theta)}(1+o(1))}).
\end{align}
It follows from (\ref{theorem2:divided integral}),
(\ref{theorem2:partI result}) and (\ref{theorem2:partIII}) that for
any $f(\vec{x_{n}})\in C_{c}(\mathbb{R}^{n})$,
\begin{align}
\label{approximate:edge}
\frac{1}{(2N^{2/3})^{n}}\int_{\mathbb{R}^{n}}
f(\vec{t_{n}})R_{n\beta}(1+\frac{t_{1}}{2N^{2/3}},\cdots,1+\frac{t_{n}}{2N^{2/3}})d\vec{t_{n}}=II'+O(N^{n}\,e^{-0.5\beta
N^{2(1-\theta)}(1+o(1))})
\end{align}
Under the transform (\ref{variable subs:edge}), we have
\begin{align}
\label{II'changed} II'=(\frac{1}{2N^{2/3}})^{n}\int_{\mathbb{R}^{n}}
\int_{1-\alpha_{_{N}}}^{1+\alpha_{_{N}}}
f\Big(u\,y_{1}+2N^{2/3}(u-1),\cdots,uy_{n}+2N^{2/3}(u-1)\Big)\nonumber\\
\times
\Psi_{N\beta}(u)R_{n\beta}^{FT}(1+\frac{y_{1}}{2N^{2/3}},\ldots,1+\frac{y_{n}}{2N^{2/3}})dudy_{1}\cdots
dy_{n}.
\end{align}
We also notice that
\begin{align}
\label{result:at the edge}
&(\frac{1}{2N^{2/3}})^{n}\int_{\mathbb{R}^{n}}
\int_{1-\alpha_{_{N}}}^{1+\alpha_{_{N}}} f(\vec{y_{n}})
\Psi_{N\beta}(u)R_{n\beta}^{FT}(1+\frac{y_{1}}{2N^{2/3}},\ldots,1+\frac{y_{n}}{2N^{2/3}})dud\vec{y_{n}}\\
&=(\frac{\pi}{2N})^{n}\int_{\mathbb{R}^{n}}
f(y_{1},\cdots,y_{n})R_{n\beta}^{FT}(1+\frac{y_{1}}{2N^{2/3}},\ldots,1+\frac{y_{n}}{2N^{2/3}})d\vec{y_{n}}(1+O(e^{-0.5\beta
N^{2(1-\theta)}(1+o(1))})).\nonumber
\end{align}
We need to prove that the difference between the right hand side of
(\ref{II'changed}) and (\ref{result:at the edge}) is zero when $N$
tends to infinity, i.e.,
\begin{align}
\label{theorem2:difference} &
\lim_{N\rightarrow\infty}\big|(\frac{1}{2N^{2/3}})^{n}\int_{\mathbb{R}^{n}}
\int_{1-\alpha_{_{N}}}^{1+\alpha_{_{N}}} [f(u\,y_{1}+2N^{2/3}(u-1),
\cdots,uy_{n}+2N^{2/3}(u-1))-f(\vec{y_{n}})]\nonumber\\&\times
\Psi_{N\beta}(u)R_{n\beta}^{FT}(1+\frac{y_{1}}{2N^{2/3}},\cdots,1+\frac{y_{n}}{2N^{2/3}})dud\vec{y_{n}}\big|=0.
\end{align}
For any $\epsilon>0$, there exists some $\delta(\epsilon)>0$ such
that $|f(\vec{x})-f(\vec{y})|<\epsilon,$ whenever
$\|\vec{x}-\vec{y}\|<\delta$. Since $1-\alpha_{_{N}}\leq
u\leq1+\alpha_{_{N}}$ and $2/3<\theta<1$, we can choose a ball
$\mathbf{B}_{R}$ such that $supp(f)\subset\mathbf{B}_{R}$ and  $$
\{(uy_{1}+2N^{2/3} (u-1),\ldots,uy_{n}+2N^{2/3}
(u-1))|\vec{y_{n}}\in supp(f),1-\alpha_{_{N}}\leq
u\leq1+\alpha_{_{N}}\}\subset\mathbf{B}_{R}.$$ For $\vec{y_{n}}\in
supp(f)$, there exist $N_{0}$ independent on $\vec{y_{n}}$ such that
\begin{align}
&\|(u\,y_{1}+2N^{2/3}(u-1),
\cdots,uy_{n}+2N^{2/3}(u-1)-(y_{1},\cdots,y_{n})\|\nonumber\\
&\leq
\sqrt{\sum_{i=1}^{n}(2N^{2/3-\theta}+|y_{i}|N^{-\theta})^{2}}\leq\sqrt{n}
(R N^{-\theta}+2N^{2/3-\theta})\leq\delta(\epsilon)
\end{align} for $N>N_{0}$. Therefore, $\forall\,\vec{y_{n}}\in supp(f)$
\begin{equation}
|f(u\,y_{1}+2N^{2/3}(u-1),
\cdots,uy_{n}+2N^{2/3}(u-1))-f(y_{n},\ldots,y_{n})|<\epsilon.
\end{equation}
Furthermore, we get
\begin{align}
\label{dominate theorem2} &|II'-(\ref{result:at the edge})|\leq
\epsilon(\frac{1}{2N^{2/3}})^{n}\int_{\mathbf{B}_{R}}
\int_{1-\alpha_{_{N}}}^{1+\alpha_{_{N}}}\,\Psi_{N\beta}(u)\,R_{n\beta}^{FT}(1+\frac{y_{1}}{2N^{2/3}},\ldots,1+\frac{y_{n}}{2N^{2/3}})dud\vec{y_{n}}\nonumber\\
&=\epsilon
\int_{1-\alpha_{_{N}}}^{1+\alpha_{_{N}}}\Psi_{N\beta}(u)du\,(\frac{1}{2N^{2/3}})^{n}
\int_{\mathbf{B}_{R}}R_{n\beta}^{FT}(1+\frac{y_{1}}{2N^{2/3}},\cdots,1+\frac{y_{n}}{2N^{2/3}})d\vec{y_{n}}\nonumber\\
&\leq 2\epsilon (\frac{1}{2N^{2/3}})^{n}
\int_{\mathbf{B}_{R}}R_{n\beta}^{FT}(1+\frac{y_{1}}{2N^{2/3}},\cdots,1+\frac{y_{n}}{2N^{2/3}})d\vec{y_{n}}\nonumber\\
& \leq 2\epsilon\,M_{_{R}}.
\end{align}
Here we have used Lemma \ref{lemma3} below. Thus,
(\ref{theorem2:difference}) can be obtained. The proof of  Lemma
\ref{lemma3} will be postponed
until  completing the proof. %According to Here we
%use (\ref{integral equation: abbrevation}). The subsequent Lemma
%\ref{proposition:upper bound about edge} means that
%(\ref{theorem2:difference}) is true. We will prove the Lemma
%\ref{proposition:upper bound about edge} after completing the proof
%of this theorem.
Combining (\ref{approximate:edge}), (\ref{II'changed}),
(\ref{result:at the edge}) and (\ref{theorem2:difference}), we have
\begin{align}
&\int_{\mathbb{R}^{n}}
f(t_{1},\cdots,t_{n})\frac{1}{(2N^{2/3})^{n}}R_{n\beta}(1+\frac{t_{1}}{2N^{2/3}},\cdots,1+\frac{t_{n}}{2N^{2/3}})dt_{1}\cdots
dt_{n}\nonumber\\
&=\int_{\mathbb{R}^{n}}
f(y_{1},\cdots,y_{n})(\frac{1}{2N^{2/3}})^{n}R_{n\beta}^{FT}(1+\frac{y_{1}}{2N^{2/3}},\cdots,1+\frac{y_{n}}{2N^{2/3}})d\vec{y_{n}}(1+o(1))+o(1),\nonumber
\end{align}
for large $N$. From (\ref{airykernel}), we complete the proof of
Theorem \ref{theoremforFT}: (ii).\end{proof}

\begin{lemma}
\label{lemma3} Let $\beta=1,2,4$, for any fixed $R>0$, there exists
a constant $M_{_{R}}$ such that
\begin{equation}
\label{uniform bounded: at the edge}
(\frac{1}{2N^{2/3}})^{n}\int_{\mathbf{B}_{R}}R_{n\beta}^{FT}(1+\frac{y_{1}}{2N^{2/3}},\cdots,1+\frac{y_{n}}{2N^{2/3}})dy_{1}\cdots
dy_{n}\leq M_{_{R}}.
\end{equation}
\end{lemma}

\begin{proof}
 Choose any positive
$\delta<R$, there exists $N_{o}(R,\delta)$ such that for $N>N_{0}$,
$\vec{y_{n}}\in\mathbf{B}_{R}$, we have
\begin{equation}
\sqrt{\sum_{i=1}^{n}(uy_{i}+2N^{2/3}(u-1))^{2}}\leq
\sqrt{\sum_{i=1}^{n}(u|y_{i}|+2N^{2/3-\theta})^{2}}<R+\delta
\end{equation}
where $u\in[1-N^{-\theta},1+N^{-\theta}]$, $2/3<\theta<1$. Let
$\eta\in(0,1)$ be a real number and let $\phi(t)$ be a smooth
decreasing function on $[0,R+\delta)$ such that $\phi(t) = 1$ for $t
\in[0,R+\delta)$ and $\phi(t)=0$ for $t\geq(1+\eta)(R+\delta)$. Set
$\varphi(\vec{x_{n}})=\phi(\|\vec{x_{n}}\|)$ for $\vec{x_{n}} =
(x_{1},\cdots, x_{n})\in\mathbb{R}^{n}$.  For $N>N_{0}$, we have
\begin{align}
\label{lemma3:scaling inequality}
(\frac{1}{2N^{2/3}})^{n}\int_{\mathbf{B}_{R}}R_{n\beta}^{FT}(1+\frac{y_{1}}{2N^{2/3}},\cdots,1+\frac{y_{n}}{2N^{2/3}})dy_{1}\cdots
dy_{n}\nonumber\\
\leq(\frac{1}{2N^{2/3}})^{n}\int_{\mathbb{R}^{n}}\varphi\Big(u\,y_{1}+2N^{2/3}(u-1),\cdots,uy_{n}+2N^{2/3}(u-1)\Big)\nonumber\\\times
R_{n\beta}^{FT}(1+\frac{y_{1}}{2N^{2/3}},\cdots,1+\frac{y_{n}}{2N^{2/3}})dy_{1}\cdots
dy_{n}.
\end{align}
Multiplying by $\Psi_{N\beta}(u)$ and then integrating
(\ref{lemma3:scaling inequality}) with respect to $u$ on
$[1-\alpha_{_{N}},1+\alpha_{_{N}}]$, one obtains
\begin{align}
&\int_{1-\alpha_{_{N}}}^{1+\alpha_{_{N}}}\Psi_{N\beta}(u)du\,(\frac{1}{2N^{2/3}})^{n}\int_{\mathbf{B}_{R}}R_{n\beta}^{FT}(1+\frac{y_{1}}{2N^{2/3}},\cdots,1+\frac{y_{n}}{2N^{2/3}})dy_{1}\cdots
dy_{n}\nonumber\\
&\leq(\frac{1}{2N^{2/3}})^{n}\int_{\mathbb{R}^{n}}
\int_{1-\alpha_{_{N}}}^{1+\alpha_{_{N}}}
\varphi\Big(u\,y_{1}+2N^{2/3}(u-1),\cdots,uy_{n}+2N^{2/3}(u-1)\Big)\nonumber\\
&\times\,
\Psi_{N\beta}(u)\,R_{n\beta}^{FT}(1+\frac{y_{1}}{2N^{2/3}},\cdots,1+\frac{y_{n}}{2N^{2/3}})dudy_{1}\cdots
dy_{n}\nonumber\\
&=\frac{1}{(2N^{2/3})^{n}}\int_{\mathbb{R}^{n}}
\varphi(\vec{t_{n}})R_{n\beta}(1+\frac{t_{1}}{2N^{2/3}},\cdots,1+\frac{t_{n}}{2N^{2/3}})d\vec{t_{n}}+O(N^{n}\,e^{-0.5\beta
N^{2(1-\theta)}(1+o(1))}).
\end{align}
Here we make use of (\ref{approximate:edge}) and (\ref{II'changed}).
Thus,
 (\ref{airykernel}) implies
 (\ref{uniform bounded: at the edge}).

%Case 2: (\ref{lemma3:condition2}) is true. Similarly, making use of
%$\varphi(\vec{y_{n}})$, we can obtain
%\begin{align}
%&(\frac{1}{2N^{2/3}})^{n}\int_{\mathbf{B}_{R}}R_{n\beta}^{FT}(1+\frac{y_{1}}{2N^{2/3}},\cdots,1+\frac{y_{n}}{2N^{2/3}})dy_{1}\cdots
%dy_{n}\nonumber\\
%%\leq(\frac{1}{2N^{2/3}})^{n}\int_{\mathbb{R}^{n}}\varphi\Big(u\,y_{1}+2N^{2/3}(u-1),\cdots,uy_{n}+2N^{2/3}(u-1)\Big)\nonumber\\
%&\leq(\frac{1}{2N^{2/3}})^{n}\int_{\mathbb{R}^{n}}\varphi(y_{1},\cdots,y_{n})
%R_{n\beta}^{FT}(1+\frac{y_{1}}{2N^{2/3}},\cdots,1+\frac{y_{n}}{2N^{2/3}})dy_{1}\cdots
%dy_{n}.
%\end{align}
Hence we complete the proof of this lemma.\end{proof}

\section{Bounded Trace Gaussian Ensembles }

 We first give a representation of correlation functions for the bounded trace Gaussian ensembles
 in terms of these for the fixed trace Gaussian ensembles, just as the authors dealt with  bounded trace Laguerre unitary ensemble in \cite{lz}.
\begin{proposition}
\label{relationsforBTLUE} Let $R_{n\beta}^{BT,r}$ and
$R_{n\beta}^{FT,r}$ be the $n$-point correlation functions for the
bounded trace and fixed trace  Gaussian ensembles respectively, then
we have the following relation
\begin{equation}
R_{n\beta}^{BT,r}(x_{1},\ldots
,x_{n})=\int_{0}^{1}N_{\beta}\,u^{N_{\beta}-1}\,
\frac{1}{u^{n}}R_{n\beta}^{FT,r}(\frac{x_{1}}{u},\ldots
,\frac{x_{n}}{u})\,du , \label{BTLUE and FTLUE}
\end{equation}
where $N_{\beta}=N+\beta N(N-1)/2$.
 \end{proposition}

\begin{proof} It suffices to prove
\begin{equation}
R_{n\beta}^{BT,r}(x_{1},\ldots
,x_{n})=\int_{0}^{r}\frac{N_{\beta}}{r^{N_{\beta}}}u^{N_{\beta}-1}\,
(\frac{r}{u})^{n}R_{n\beta}^{FT,r}(\frac{r}{u}x_{1},\ldots
,\frac{r}{u}x_{n})\,du.
\end{equation}
For every $u>0$, let

 \begin{equation}
\Omega_{N}(u)=\{(x_{1},\ldots,x_{N})|
\sum_{j=1}^{N}x^{2}_{j}=u^{2}\}
\end{equation}
be the sphere in $\mathbb{R}^{N}$, which carries the volume element
induced by the standard Euclidean metric on $\mathbb{R}^{N}$,
denoted by $u^{N-1}d \,\sigma_{N}$. For $h\in
L^{\infty}(\mathbb{R}^{N})$, let $<h(\cdot)>_{\theta}$ and
$<h(\cdot)>_{\delta}$ denote  the ensemble average taken in the
bounded trace and fixed trace ensembles, respectively. From
(\ref{fixedtracepdf}) and (\ref{boundedtracepdf}), we have
\begin{flalign*}
&<h(\cdot)>_{\theta}\\
&= \frac{1}{Z_{N\beta}^{BT,r}}\int_{0}^{r}u^{N-1}d\,u
\int_{\Omega_{N}(u)}h(x_{1},\ldots,x_{N})\prod_{1\leq i<j\leq
N}|x_{i}-x_{j}|^{\beta}d \,\sigma_{N}\\
&= \frac{1}{Z_{N\beta}^{BT,r}}\int_{0}^{r}u^{N_{\beta}-1}d\,u
\int_{\Omega_{N}(1)}h(u x_{1},\ldots,u x_{N})\prod_{1\leq i<j\leq
N}|x_{i}-x_{j}|^{\beta}d \,\sigma_{N},
\end{flalign*}
and
\begin{flalign*}
&<h(a\,\cdot)>_{\delta}\\
&=\frac{1}{Z_{N\beta}^{FT,r}} \int_{\Omega_{N}(r)}r^{N-1}\,h(a
x_{1},\ldots,a x_{N})\prod_{1\leq i<j\leq
N}|x_{i}-x_{j}|^{\beta}d \,\sigma_{N}\\
&=\frac{1}{Z_{N\beta}^{FT,r}}
\int_{\Omega_{N}(1)}r^{N_{\beta}-1}\,h(a r x_{1},\ldots,a r
x_{N})\prod_{1\leq i<j\leq N}|x_{i}-x_{j}|^{\beta}d \,\sigma_{N}.
\end{flalign*}
Choose $a=\frac{u}{r}$, we get
\begin{equation}
<h(\cdot)>_{\theta} =\frac{Z_{N\beta}^{FT,r}} {Z_{N\beta}^{BT,r}}
\int_{0}^{r}(\frac{u}{r})^{N_{\beta}-1}<h(\frac{u}{r}\,\cdot)>_{\delta}
d\,u.
\end{equation}
Setting $h\equiv 1$, we get the ratio of the partition functions
$Z_{N\beta}^{FT,r}$ and $Z_{N\beta}^{BT,r}$. Substituting this
ratio, we then obtain
\begin{equation}
<h(\cdot)>_{\theta}= \int_{0}^{r}\frac{N_{\beta}}{r^{N_{\beta}}} u^{
N_{\beta}-1 } <h(\frac{u}{r}\,\cdot)>_{\delta} d\,u.
\end{equation}
In particular, taking
\begin{equation}
h(x_{1},\ldots,x_{N})=\sum_{1\leq i_{1}<\cdots <i_{n}\leq N
}f(x_{i_{1}},\ldots,x_{i_{n}}),
\end{equation}
we have
\begin{flalign*}
&\int_{\mathbb{R}^{n}}f(x_{1},\ldots,x_{n})R_{n\beta}^{BT,r}(x_{1},\ldots,x_{n})\,d^{n}x\nonumber \\
&=\int_{0}^{r}\frac{N_{\beta}}{r^{N_{\beta}}} u^{ N_{\beta}-1 }d\,u
\int_{\mathbb{R}^{n}}f(\frac{u}{r}x_{1},\ldots,\frac{u}{r}x_{n})\,
R_{n\beta}^{FT,r}(x_{1},\ldots,x_{n})\,d^{n}x\nonumber \\
&=\int_{\mathbb{R}^{n}}f(x_{1},\ldots,x_{n})\,d^{n}x
\int_{0}^{r}\frac{N_{\beta}}{r^{N_{\beta}}}u^{N_{\beta}-1}\,
(\frac{r}{u})^{n}R_{n\beta}^{FT,r}(\frac{r}{u}x_{1},\ldots
,\frac{r}{u}x_{n})\,du.
\end{flalign*}
Since $R_{n\beta}^{BT,r}$ and $R_{n\beta}^{FT,r}$ are both
continuous, we complete the proof.
\end{proof}

Next,   we notice a ``sharp" concentration phenomenon along the
radial coordinate between correlation functions of the the bounded
trace and fixed trace  Gaussian ensembles. Although its proof  is
simple, the following lemma plays a crucial role in dealing with
local statistical properties of the eigenvalues between the fixed
and bounded ensembles.

\begin{lemma}
\label{concentrationlemma2}Let $\{b_{N}\}$ be a sequence such that
$b_{N}\rightarrow 0$ but $N^{2} b_{N}\rightarrow\infty$ as
$N\rightarrow\infty$, then we have
\begin{equation}\int_{0}^{1}N_{\beta}\,u^{N_{\beta}-1}\,
\,du=\int_{u_{-}}^{1} N_{\beta}\,u^{N_{\beta}-1}\,du+ e^{-0.5 \beta
N^{2} b_{N}(1+o(1))},
\end{equation}
 where
$u_{-}=1-b_{N}$.
 \end{lemma}
\begin{proof}
\begin{flalign*}
\int_{0}^{u_{-}} N_{\beta}\,u^{N_{\beta}-1}
\,du&=(1-b_{N})^{N_{\beta}}=e^{N_{\beta}\ln
(1-b_{N})}\\&=e^{N_{\beta}\big(-b_{N}+O(b^{2}_{N})\big)}=e^{-0.5
\beta N^{2} b_{N}(1+o(1))}.
\end{flalign*}
This completes the proof.
\end{proof}

\begin{remark}
In Lemma \ref{concentrationlemma2}, let us take $b_{N}=N^{-\kappa},
\kappa\in (0,2)$. Since the ``rate" index $\kappa$ can be chosen
larger than 1 while the scaling in the bulk is proportional to
$N^{-1}$ and at the soft edge of the spectrum is proportional to
$N^{-2/3}$, in principle  we can prove all local statistical
properties of the eigenvalues between  the fixed and bounded trace
Gaussian ensembles are identical in the limit. Such arguments also
apply to  the equivalence of ensembles between the fixed trace and
bounded trace ensembles with monomial potentials, where we exploit
some homogeneity of the monomial potentials. In addition, we notice
that equivalence of $n$-point resolvents of the   fixed and bounded
trace ensembles with monomial potentials has turned out to be
identical in the limit in \cite{acmv2}.
 %Such arguments apply to  the
%equivalence of ensembles between the fixed trace and bounded trace
%ensembles with monomial potentials, where we exploit some
%homogeneity of the monomial potentials.
\end{remark}

Now we turn to the proof of Theorem \ref{theoremforBT}, by using the
associated results about the fixed trace ensembles.

\begin{proof} The proof is very similar to that of Theorem \ref{theoremforFT}, we only point
out some different places in the bulk case for bounded trace GUE.

 In Lemma \ref{concentrationlemma2}
, choose $b_{N}=N^{-\kappa}, \,\kappa \in (1,2)$. The change of
variables corresponding to (\ref{variable subs:edge}) reads:
\begin{equation}
t_{i}=(u-1)N\, x \omega(x)+u y_{i},i=1,\ldots ,n
\end{equation}
where fixed $x\in (-1,1)$. The condition that $b_{N}=N^{-\kappa},
\,\kappa
 \in (1,2)$ ensures $(1-u)N\leq  N^{-\kappa+1}\rightarrow 0$ as
 $N\longrightarrow \infty$ for $u\in [u_{-},1]$. On the other hand, by (\ref{sinekernelforFTGUE}), the following fact similar to
 Lemma \ref{lemma3} is obvious:
 for any fixed $R>0$,
\begin{equation}
\frac{1}{(N\omega(x))^{n}}\int_{B_{R}}R_{n\beta}^{FT}(x+
\frac{y_{1}}{N\omega(x)},\cdots
,x+\frac{y_{n}}{N\omega(x)})d^{n}y\leq C_{R}.
\end{equation}
Here  $B_{R}$ is the ball of the
radius $R$ in $\mathbb{R}^{n}$ centered at zero, and  $%
C_{R}$ is a constant. Using Proposition \ref{relationsforBTLUE} and
universality of sine-kernel (\ref{sinekernelforFTGUE}) for fixed
trace GUE, we complete the proof after a similar procedure.
\end{proof}

\section*{Acknowledgments}
D.-Z. Liu and D.-S. Zhou respectively thank Prof. Zheng-Dong Wang
and Prof. Tao Qian for their encouragement and support. The work of
D.-S. Zhou is supported by research grant of the University of Macau
No. FDCT014/2008/A1.

\end{document}